\documentclass[preprint,aps,nofootinbib]{revtex4}

\usepackage{graphicx}
\usepackage{dcolumn}
\usepackage{bm}
\usepackage{epsfig}
\usepackage{graphicx}
\usepackage{float}
\usepackage[caption=false]{subfig}

\usepackage{amsmath}
\usepackage{amsfonts}
\usepackage{amssymb}

\usepackage{amsmath}
\usepackage{amsfonts}
\usepackage{amssymb}
\usepackage{color}

\def\be{\begin{equation}}
\def\ee{\end{equation}}
\def\bea{\begin{eqnarray}}
\def\eea{\end{eqnarray}}

\def\ss2l{SS2$\ell$}
\def\3l{3$\ell$}

\def\slashchar#1{\setbox0=\hbox{$#1$}           
   \dimen0=\wd0                                 
   \setbox1=\hbox{/} \dimen1=\wd1               
   \ifdim\dimen0>\dimen1                        
      \rlap{\hbox to \dimen0{\hfil/\hfil}}      
      #1                                        
   \else                                        
      \rlap{\hbox to \dimen1{\hfil$#1$\hfil}}   
      /                                         
   \fi}

\begin{document}

\title{Same-Sign Dilepton Excesses and Vector-like Quarks}
\vspace*{1cm}

\author{\vspace{1cm} Chuan-Ren Chen$^{\, a}$, Hsin-Chia Cheng$^{\, b}$ and   Ian Low$^{\, c,d}$ }

\affiliation{
\vspace*{.5cm}
  \mbox{$^a$Department of Physics, National Taiwan Normal University, Taipei 116, Taiwan}\\
  \mbox{$^b$Department of Physics, University of California, Davis, CA 95616, USA}\\
\mbox{$^c$ Department of Physics and Astronomy, Northwestern University, Evanston, IL 60208, USA} \\
\mbox{$^d$ High Energy Physics Division, Argonne National Laboratory, Argonne, IL 60439, USA}\\
\vspace*{1cm}}

\begin{abstract}
Multiple analyses from  ATLAS and CMS collaborations, including searches for ttH production, supersymmetric particles and vector-like quarks,   observed excesses in the same-sign dilepton channel containing $b$-jets and missing transverse energy in the LHC Run 1 data.  In the context of little Higgs theories with T parity, we explain these excesses using vector-like T-odd quarks decaying into a top quark, a $W$ boson and the lightest T-odd particle (LTP). For heavy vector-like quarks, decay topologies containing the LTP have not been searched for at the LHC. The bounds on the masses of the T-odd quarks can be estimated in a simplified model approach by adapting the search limits for top/bottom squarks in supersymmetry. Assuming a realistic decay branching fraction, a benchmark with a 750 GeV T-odd $b^\prime$ quark is proposed. We also comment on the possibility to fit excesses in different analyses in a common framework.

\end{abstract}


\maketitle

\section{Introduction}
\label{sec:intro}

It was recently pointed out  that both ATLAS and CMS observed in their  Run 1 data mild excesses in the same-sign dilepton (SS2l) channel, which contains final states with same-sign dilepton, $b$-jets, and missing transverse energy ($E_{\rm T}^{\rm miss}$) \cite{Huang:2015fba}. The excesses were observed in five semi-independent analyses, three from the ATLAS \cite{Aad:2014pda,Aad:2015gdg,atlastth} and two from the CMS \cite{Khachatryan:2014qaa,Chatrchyan:2013fea}, using different background subtraction methods. In particular, the ATLAS exotica search \cite{Aad:2015gdg} and the CMS ttH Higgs search \cite{Khachatryan:2014qaa}  reported about  $2\sigma$ and $2.5\sigma$ significance, respectively, for the excesses. More recently, measurements on the standard model (SM) process ${\rm t\bar{t}W}$ from ATLAS \cite{Aad:2015eua} and CMS \cite{Khachatryan:2015sha} have also reported seeing excesses above the SM expectation in the SS2l channel. 

A supersymmetric interpretation of the SS2l excess was put forward in Ref.~\cite{Huang:2015fba}, which proposed using top squarks (stops) or bottom squarks (sbottoms) decaying into two top quarks, two $W$ bosons, and $E_{\rm T}^{\rm miss}$ to explain the excess. However, it is worth recalling that SS2l, $b$-jets and $E_{\rm T}^{\rm miss}$ are generic signatures of many beyond-the-SM theories and, therefore, not unique to supersymmetry. In fact, it is well-known that many scenarios could ``fake" supersymmetry at the LHC, due to the complex environment of $pp$ collision \cite{Cheng:2002ab}. For example,  decays  of the stop could be mimicked by that of a fermionic top partner, unless one can measure the spin of the new particles involved in the decay chain, which is a challenging measurement at the LHC \cite{Wang:2008sw}.  Indeed, should the SS2l excess be confirmed at the LHC Run 2, it would be of paramount importance to determine the specific quantum number of the new particles associated with the excess.

In this study we would like to proceed in an exploratory spirit that is appropriate for this nascent subject, and consider alternative possibilities, other than supersymmetry, for the SS2l excess. The simplest possibility is to invoke a new conserved quantum number at the TeV scale, the T parity, under which the SM particles are neutral and the new particles are charged \cite{Cheng:2003ju}.  Collider phenomenology of T parity is very similar to that of  R parity in supersymmetry \cite{Martin:1997ns} and KK parity in Universal Extra Dimensions \cite{Appelquist:2000nn}, in that  all T-odd particles can only be pair-produced and subsequently cascade-decay into SM particles plus the lightest T-odd particle (LTP), which carries away extra $E_{\rm T}^{\rm miss}$ in collider detectors.

This work is organized as follows. In Section \ref{sec:Tparity} we give a brief overview of T parity and an estimate on the collider bound on the mass of T-odd quarks, followed by Section \ref{sect:ss2llhc} which discusses fitting the SS2l excess in the context of SM ttH searches at the LHC Run 1 and beyond. In Section \ref{sect:broadpic} we consider whether it is possible to explain using a common benchmark  the excesses observed by analyses outside of the SM ttH searches. Then we provide an outlook in Section \ref{sect:outlook}.

\section{A Simplified Overview of T parity}
\label{sec:Tparity}

In this section we provide a ``simplified" overview of T parity and consider collider bounds on T-odd quarks.

\subsection{Simplified T Parity}

The case for  a new symmetry at the TeV scale, under which all SM particles are neutral, is motivated by two considerations \cite{Cheng:2003ju}:

\begin{itemize}

\item Precision Electroweak Constraints: If we parameterize flavor-conserving new physics  in terms of higher-dimensional operators, precision electroweak constraints indicate that the mass scale suppressing these operators tend to be at around 5 -- 10 TeV, assuming all dimensionless coefficients to be order unity \cite{Barbieri:1999tm}. This is referred to as the little hierarchy problem, since naturalness principle expects new physics at around 1 TeV scale to stabilize the Higgs mass. If there exists a new symmetry at the TeV scale such that all SM particles are neutral while all new particles are charged under the new symmetry, then new particles would enter into precision electroweak observables only through loop-induced effects, thereby introducing a loop factor of about $1/16\pi^2$ in front of the higher dimensional operators. This allows one to lower the scale of new physics down to below 1 TeV.

\item Dark Matter: The existence of dark matter calls for (at least) a new particle that is electrically neutral and stable at the cosmological time scale. A simple possibility is to make the dark matter particle absolutely stable. This can be achieved again by postulating a new symmetry such that all SM particles are neutral. Then the lightest particle charged under the new symmetry will be stable, assuming the symmetry is exact, or almost exact so that the particle is stable cosmologically. If such a particle is also electrically neutral, it can be a dark matter candidate. 

\end{itemize}

There are many realizations of such a new TeV symmetry in explicit models. The most popular models use the simplest symmetry: a new parity ($Z_2$) symmetry. R parity in supersymmetry \cite{Martin:1997ns}, KK parity in Universal Extra Dimensions \cite{Appelquist:2000nn} and warped extra dimension \cite{Randall:1999ee, Agashe:2007jb}, and T parity in little Higgs models \cite{ArkaniHamed:2001nc,Cheng:2003kk} all fall into this simple category. Nevertheless, larger symmetry groups such as the $Z_3$ group have also been employed \cite{Agashe:2004ci}. 

It is possible to adopt the ``simplified model" approach~\cite{Alves:2011wf}, by postulating the existence of a new parity, without committing to a specific model realization. In particular, one could assume there is a ``simplified T parity," under which all SM particles are neutral and all new particles are charged. Then at colliders any new particles must be pair-produced, in order to conserve the simplified T parity, and eventually  cascade-decay into the LTP. The LTP would serve as the dark matter candidate, assuming it is electrically neutral, and manifest itself as $E_{\rm T}^{\rm miss}$ in colliders. Phenomenology of the simplified T parity at the LHC is very similar to that of supersymmetry with R parity, which is characterized by leptons, jets and $E_{\rm T}^{\rm miss}$.\footnote{For a related approach to the simplified T parity, see Ref.~\cite{Anandakrishnan:2015yfa}.}

In the end, many collider signatures of R parity conserving SUSY can  be mimicked by simplified T parity, which allows more freedom in choosing the spin quantum number of the mother and intermediate particles in the decay chain. For example, decays of bottom squarks in SUSY,
\be
\label{eq:sbottom}
\tilde{b}_1 \to t + (\tilde{\chi}^-_1 \to W^- + \tilde{\chi}_1^0) \ ,
\ee
can be impersonated in simplified T parity by a T-odd vector-like quark with the same quantum number as the bottom quark:
\be
\label{eq:bprime}
b^\prime \to t + ( W_H^{-} \to W^- + A_H )\ ,
\ee
where $W_H^\pm$ is a pair of heavy charged vector boson and $A_H$ is a heavy neutral vector boson that is the LTP.  For the SS2l excess, the bottom squark decay chain Eq.~(\ref{eq:sbottom}) was discussed to some extent in Ref.~\cite{Huang:2015fba}, which focused on supersymmetric theories. In this work we consider the alternative decay chain in Eq.~(\ref{eq:bprime}) based on the (simplified) T parity. While both models can produce SS2l events, the kinematic distributions in general are different, and might be used to distinguish different models. Of course, to fully identify the decay chain requires measuring the spin of the mother and intermediate particles \cite{Wang:2006hk}, which would be a top priority should the excess be confirmed at the LHC.

Explicit constructions of the T-odd top and bottom partners in little Higgs models were given in Ref.~\cite{Cheng:2003kk,Cheng:2005as,Hubisz:2005tx}. It was shown that in order to cut off the contributions to the standard model four-fermion interactions from the Goldstone boson loop, there should be a vector-like T-odd doublet partner for every standard model fermionic doublet. In these models the branching fraction of the desired decay chain in Eq.~(\ref{eq:bprime}) is typically not 100\%, in contrast to the simplified model approach. Because the dominant contributions to the masses of the T-odd vector bosons are $SU(2)_L\times U(1)_Y$ preserving, the corresponding mixing between the T-odd partners of the neutral gauge bosons due to the electroweak breaking is small. As a result, $A_H$ is mostly the partner of the Standard Model hypercharge gauge boson $B_\mu$ and $Z_H$ is mostly the partner of the $SU(2)_L$ gauge boson $W_\mu^3$. In this case, the decay branching fractions of the T-odd $b^\prime$ and $t^\prime$ are more or less determined by the Goldstone equivalence theorem \cite{Belyaev:2006jh}:
\bea
Br(b^\prime\to tW_H) : Br(b^\prime\to bZ_H)& \approx& 2:1  \nonumber \\
Br(t^\prime\to bW_H) : Br(t^\prime\to tZ_H) &\approx& 2:1 \nonumber \ .
\eea  
The other decay channels, $b^\prime \to b A_H$ and $t^\prime \to t A_H$, are subdominant.
These observations will be taken into account when we consider model-specific T-odd quark decays.\footnote{For collider signatures of the first and second generation T-odd quarks, see Ref.~\cite{Carena:2006jx}.}

\subsection{Collider Bounds on Third Generation T-odd Quarks}

Although there are many searches for stops and sbottoms in SUSY at the LHC Run~1, there hasn't been many dedicated searches on the closely related decay chains involving T-odd quarks.\footnote{There are many collider searches for vector-like fermionic top partners. However, typically T parity was not assumed in these searches and there is no LTP carrying away additional $E_{\rm T}^{\rm miss}$ in the final states.} In particular, the stop and sbottom searches usually adopt the simplified model approach, by assuming 100\% decay branching ratio (BR)  into the final states being searched for. Current limits from the LHC Run 1 data on the stop and bottoms  are typically around 500 -- 700 GeV \cite{Aad:2015pfx}. 

Since the decays of T-odd quarks often give the same final state signatures as the squarks, in this subsection we will provide a rough estimate on the bounds on the masses of T-odd quarks, based on the experimental searches for third generation squarks. Recasting the search limits on squarks is not a straightforward task, as the signal selection efficiencies  depend non-trivially on the kinematics of the decay products \cite{Low:2013aza}, which in turn is determined by the mass spectrum of the mother and daughter particles. For example, the strong limits on the sbottom mass in the tripleton channel of the decay chain in Eq.~(\ref{eq:sbottom}) varies significantly when the lightest neutralino [which is assumed to be the lightest supersymmetric particle (LSP)] and chargino masses change \cite{cmstrilepton}. Moreover, the bound disappears completely when the LSP becomes heavier than 170 -- 240 GeV, depending on the chargino mass. It is clear that 
a full-fledge study on the experimental constraints of T-odd quarks requires dedicated efforts and is beyond the scope of this paper.

\begin{figure}[t]
\subfloat[]{\includegraphics[width=3.35in, angle=0]{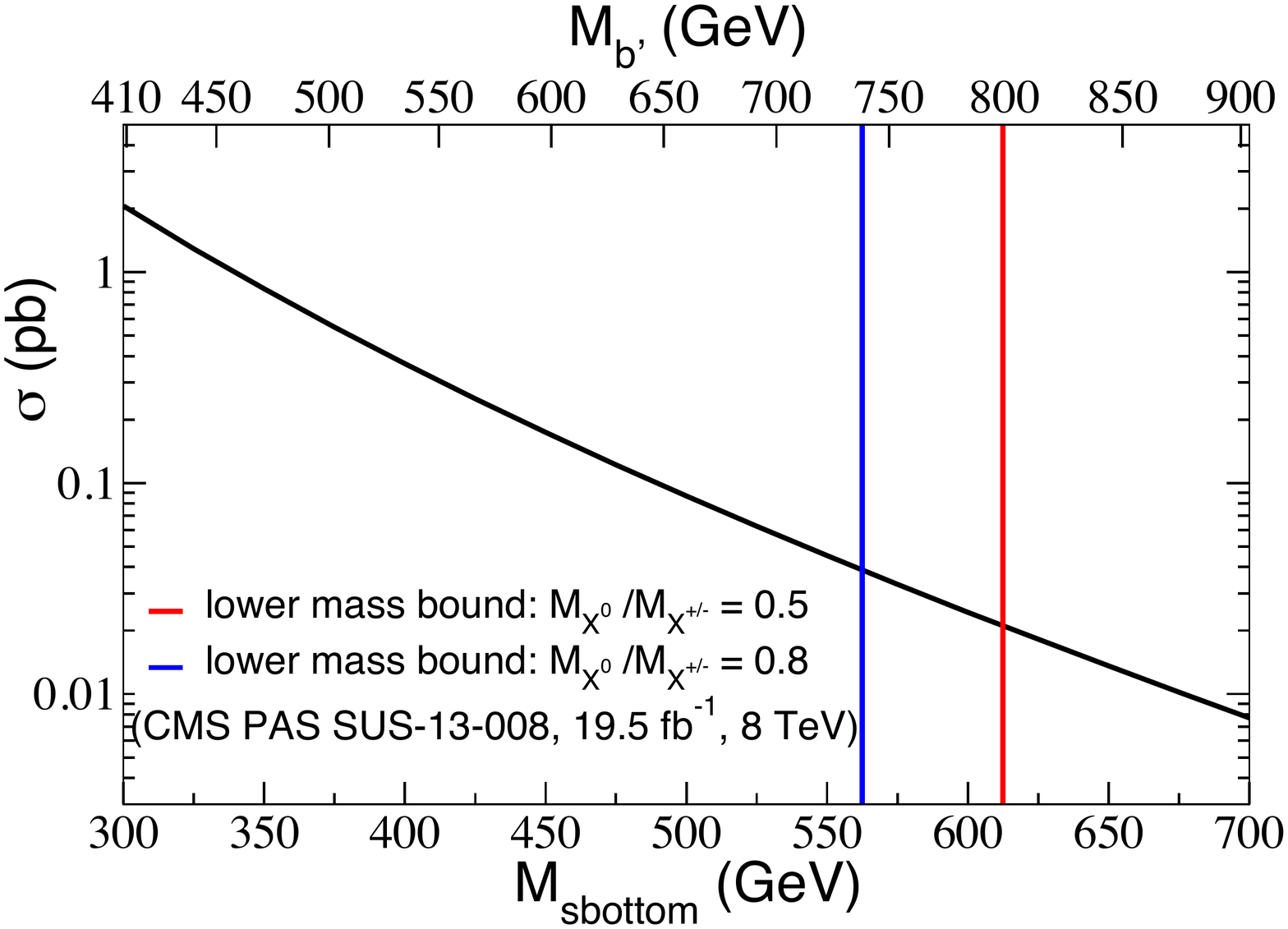}}
\subfloat[]{\includegraphics[width=3.35in, angle=0]{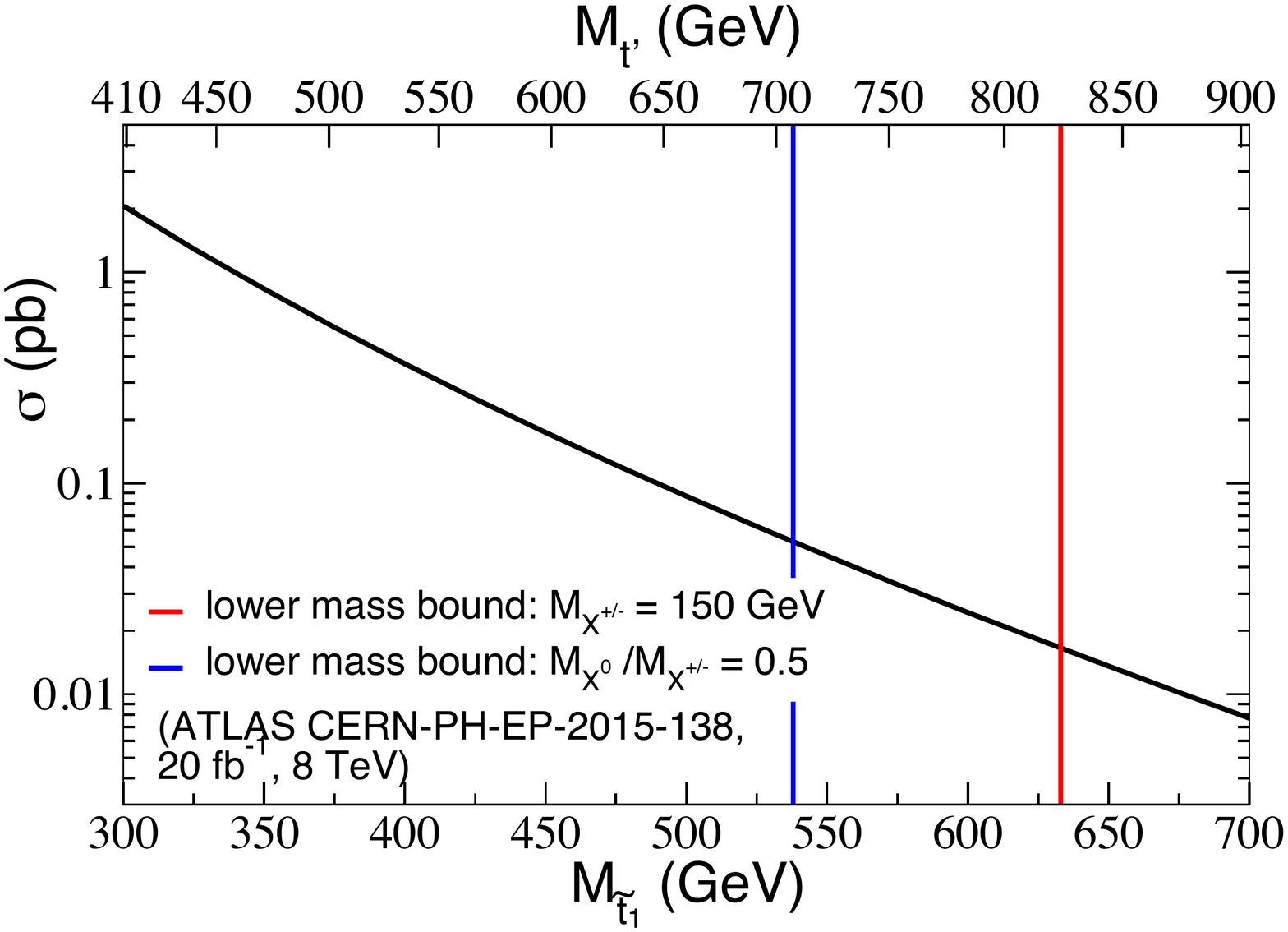}}
    \caption{\label{fig:1}\em Upper bounds on the production cross-section of third generation squarks and T-odd quarks at the 8 TeV LHC.  The bounds on the sbottom masses are obtained from the trilepton channel in Ref.~\cite{cmstrilepton}, which gives the most stringent limits, while the bounds on the stop are from Ref.~\cite{Aad:2015pfx}. In this plot the decay of squarks and T-odd quarks into the desired final states are assumed to be 100\%. }
\end{figure}

In this work we will settle for a na\"{i}ve estimate on the experimental bounds on third generation T-odd quarks by translating the {\em strongest\/} limits on the third generation squark masses into upper bounds on the production cross-sections, and then computing the $b^\prime$ and $t^\prime$ masses that give rise to the same production cross-section as the upper bounds. The outcome based on the simplified model assumption is shown in Fig.~\ref{fig:1}, where the squark and T-odd quark cross sections are quoted from the LHC SUSY Cross-section Working Group \cite{Kramer:2012bx,Borschensky:2014cia} and from {\tt HATHOR} \cite{Moch:2008ai,Langenfeld:2009tc}, respectively. We see that the lower bound on the $b^\prime$ and $t^\prime$ quarks are about 800 GeV and 825 GeV, respectively. The bounds in Fig.~\ref{fig:1} are conservative in the sense that they correspond to the most stringent limits on the $m_{\rm LSP}$ -- $m_{\tilde{b}_1/{\tilde{t}_1}}$ plane, which would loosen when the $m_{\rm LSP}$ becomes larger.

\begin{figure}[t]
\includegraphics[width=3.35in, angle=0]{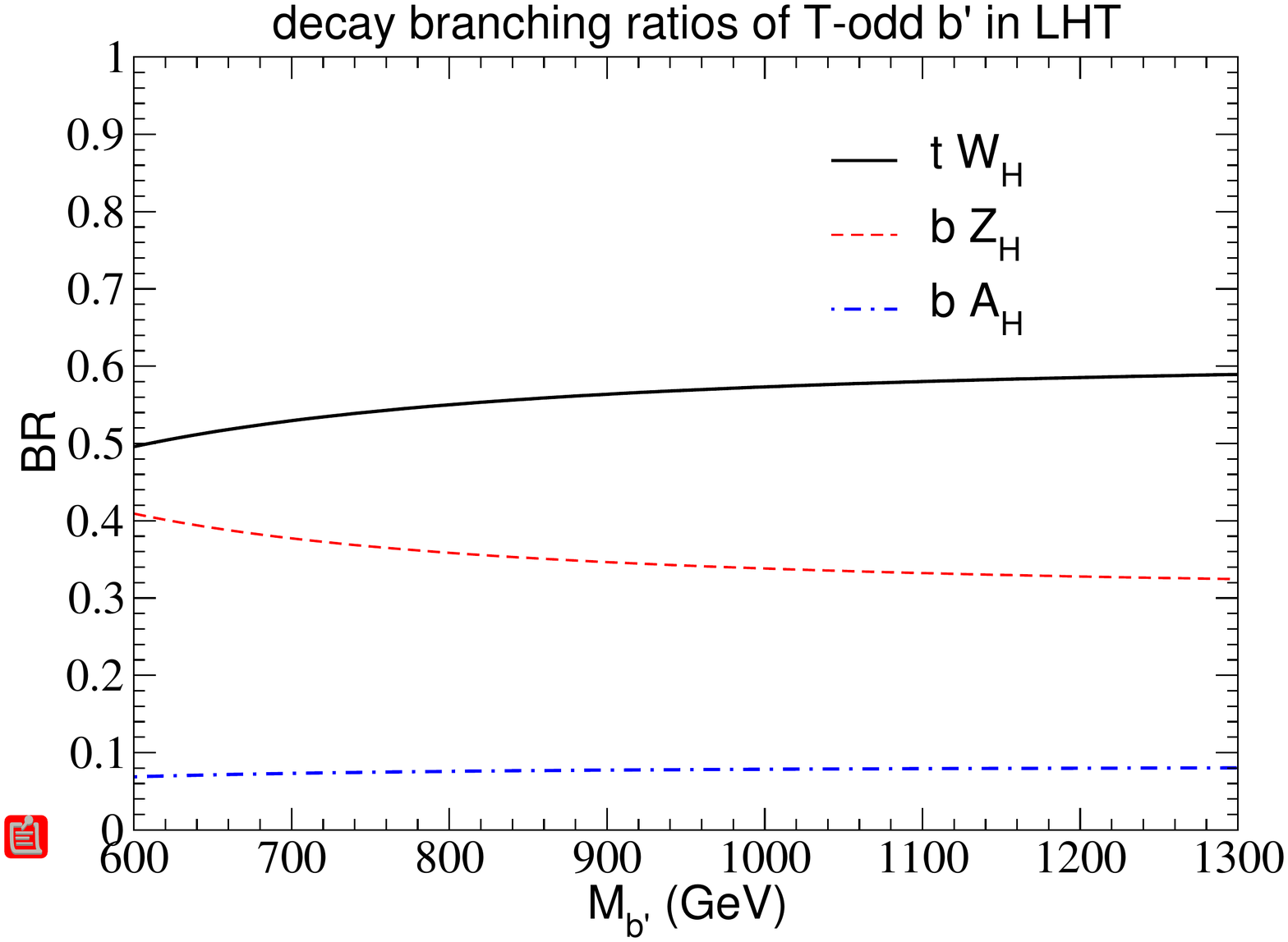}    \caption{\label{fig:2}\em Decay BR of T-odd $b^\prime$ quark in the littlest Higgs theory with T parity. As the $b^\prime$ becomes heavy, the decay BR into $bZ_H$ gets close to 1/2 of the BR into $tW_H$, as predicted by the Goldstone equivalence theorem.  The remaining decay BR to the $bA_H$ final state is quite small.}
\end{figure}

\begin{figure}[t]
\subfloat[]{\includegraphics[width=3.35in, angle=0]{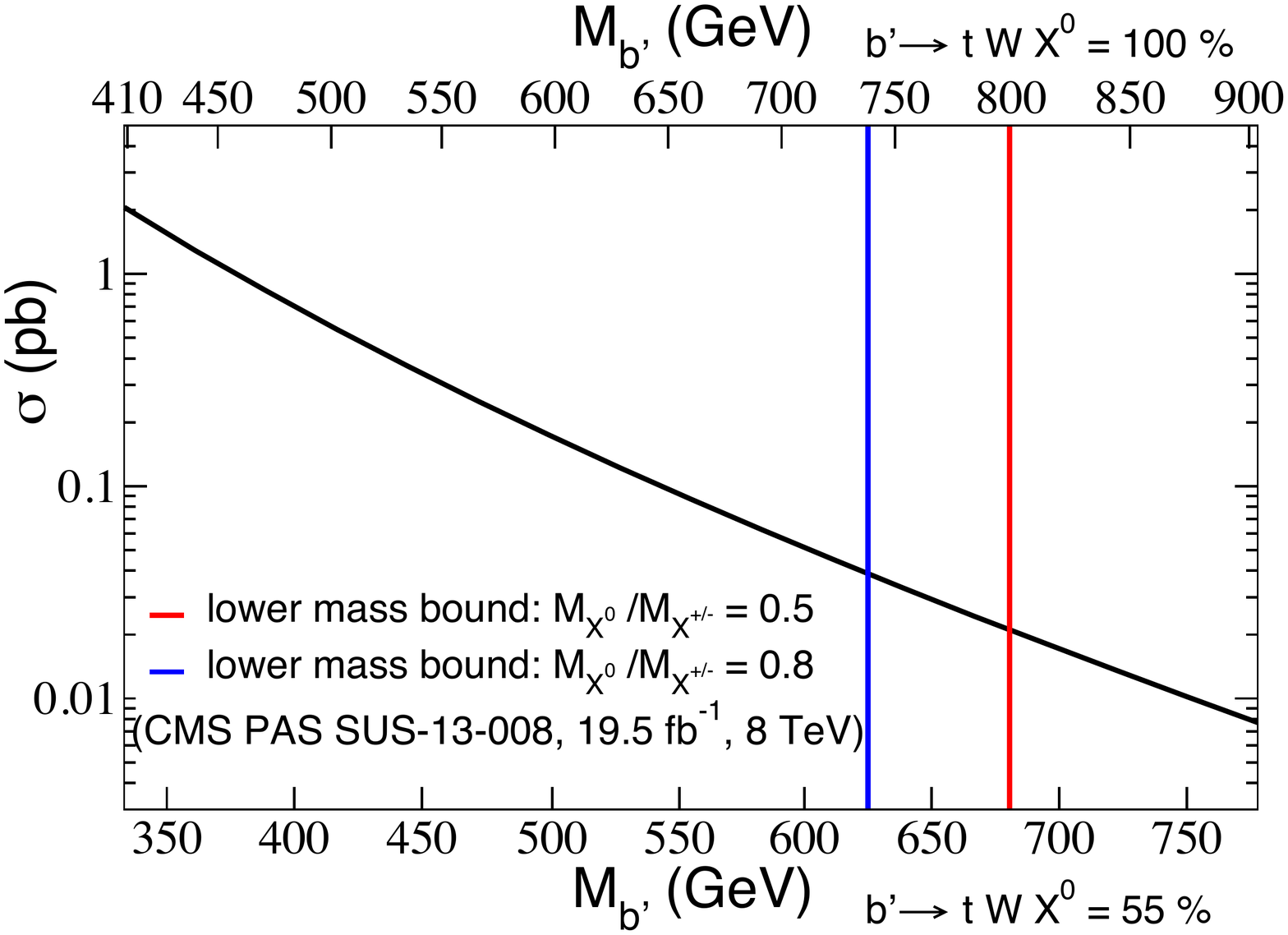}}
\subfloat[]{\includegraphics[width=3.35in, angle=0]{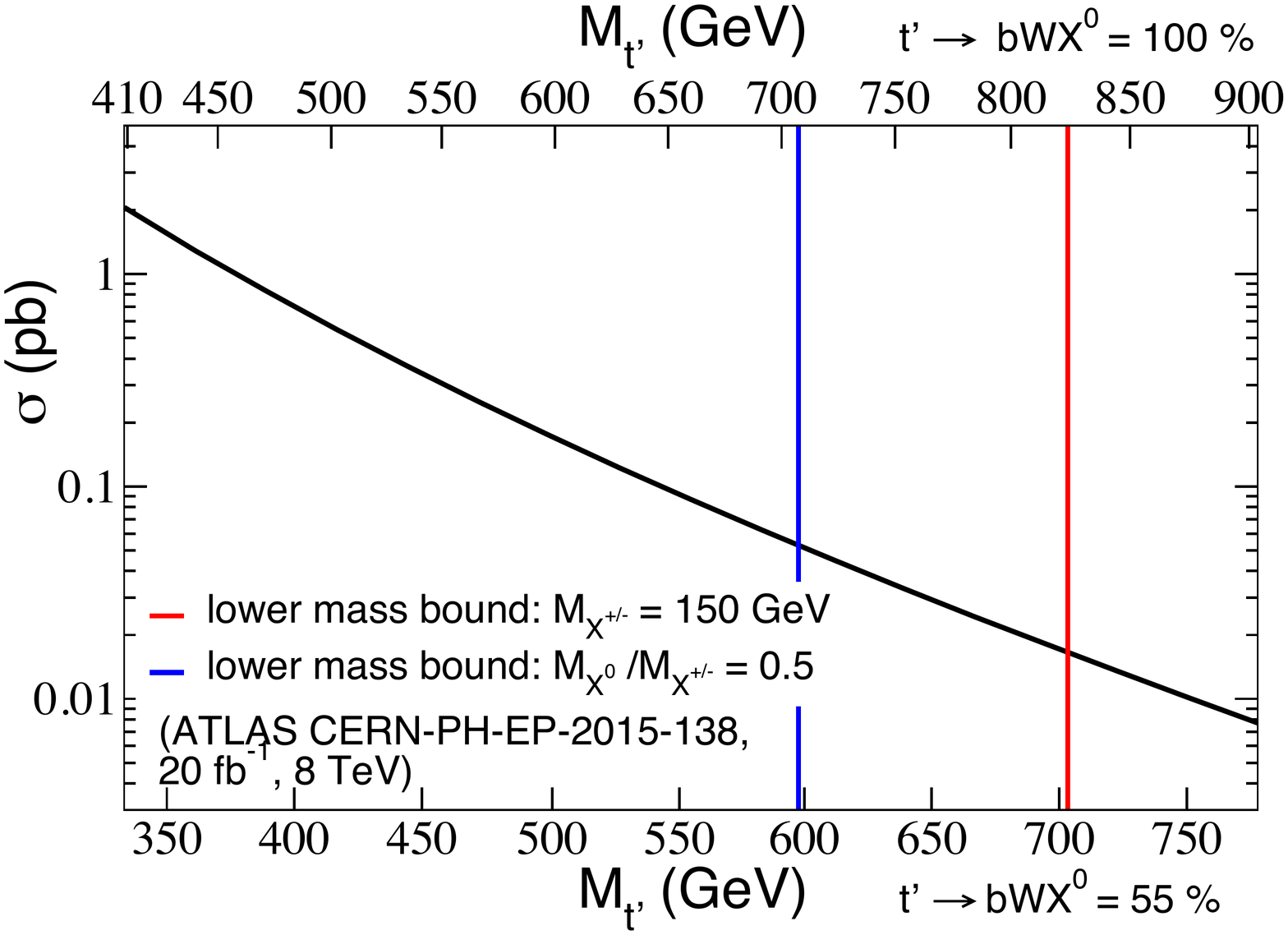}}
    \caption{\label{fig:3}\em Upper bounds on the production cross-section of third generation T-odd quarks at the 8 TeV LHC, assuming 100\% BR versus 55\% BR. }
\end{figure}

Beyond the simplified model approach, the exclusion limit degrades quickly as the decay BR decreases from the assumed 100\%, because the signal strength is usually proportional to the square of the BR due to the assumption of pair-production of the mother particles. For example, in the littlest Higgs with T parity model (LHT)~\cite{Cheng:2003kk,Belyaev:2006jh}, the decay BR of T-odd $b^\prime$ into $tW_H$ final state is typically  about 55\% at around 800 GeV, as shown in Fig.~\ref{fig:2}.  Consequently, the collider bounds on the T-odd quarks are considerably weaker in a full model than in the simplified model approach. This comparison is shown in Fig.~\ref{fig:3}, where we compared the bound assuming 100\% BR versus 55\% for both T-odd $b^\prime$ and $t^\prime$. The bounds on the $b^\prime$ ($t^\prime$) mass is only about 680 (700) GeV.

\section{Same-Sign Dilepton Signals at the LHC}
\label{sect:ss2llhc}

Having estimated the LHC bounds on third generation T-odd quarks, in this Section we consider an interpretation of the SS2l excess from the pair-production of T-odd quarks $b^\prime$. In particular, the decay chain we concentrate on is
\be
pp \to b^\prime \bar{b}^\prime \to (tW^-_H)(\bar{t}W^+_H) \to  (tW^-A_H) (\bar{t}W^+A_H) \ ,
\ee
which gives $2b+4W+E_{\rm T}^{\rm miss}$ final states and contributes to the SS2l signal region.

Following the strategy in Ref.~\cite{Huang:2015fba}, we base our numerical simulations on the selection cuts implemented in the CMS ttH analysis in Ref.~\cite{Khachatryan:2014qaa} and then normalize the $b^\prime$ signal strength to the SM ttH signal strength. More specifically,  we generate  both the T-odd $b^\prime$ pair production and  SM ttH with {\tt Madgraph/MadEvent}~\cite{Alwall:2007st}, pass the events through {\tt Pythia}~\cite{Sjostrand:2006za} for showering and then to {\tt Delphes}~\cite{deFavereau:2013fsa} for detector simulations.  The particular benchmark we study has the mass spectrum motivated by the LHT model~\cite{Cheng:2003kk,Belyaev:2006jh}:
\be
m_{b^\prime} = 750 {\rm \ GeV} \ , \qquad m_{W_H} = 320 {\rm \ GeV} \ , \qquad m_{A_H} = 66 {\rm \ GeV}\ .
\ee
The production cross section of the 750 GeV $b'$ pair at the 8 TeV LHC is 34.1 fb.
The relevant decay branching fraction is BR$(b^\prime \to tWA_H)\approx 55$\%. The selected events are required to have exactly two same-sign leptons, at least 4 jets among which two are $b$-jets.  Following Ref.~\cite{Khachatryan:2014qaa} we impose further the following kinematic cuts:
\be
\label{eq:cmscuts}
p^{\ell}_T > 20 ~{\rm GeV}\ , \quad p^j_T > 25 ~{\rm GeV}\ ,\quad {\rm L_D}> 30~{\rm GeV}\ , \quad S_T >100~{\rm GeV}\ ,
\ee
where $S_T=p^{\ell_1}_T+p^{\ell_2}_T+E_{\rm T}^{\rm miss}$ is the scalar sum of transverse momentum of two charged leptons and $E_{\rm T}^{\rm miss}$ and $\rm L_D=0.6\times E_{\rm T}^{\rm miss} + 0.4\times H^{miss}_T$ with $\rm H^{miss}_T$ being the negative vector $p_T$ sum of selected jets and two same-sign leptons. Numbers of after-the-cut events from $b^\prime$ and SM ttH are then used to calculate the ratio of signal strength $\mu_{b^\prime}/\mu_{\rm ttH}$. In the end, the total signal strength is
\be
\mu= \mu_{b^\prime} + \mu_{\rm ttH} = 2.0 \ ,
\ee
which is to be compared with the ATLAS result $\mu = 2.8_{-1.9}^{+2.1}$~\cite{atlastth} and the CMS fit $\mu = 5.3_{-1.8}^{+2.1}$~\cite{Khachatryan:2014qaa}. If we consider the ``simplified T parity model," namely the 100\% branching fraction of $b^\prime \to t W_H$, a  $b^\prime$ with a mass of $850$ GeV and production cross-section 12.85 fb will generate the similar signal strength.

\begin{figure}[t]
\subfloat[]{\includegraphics[width=3.35in, angle=0]{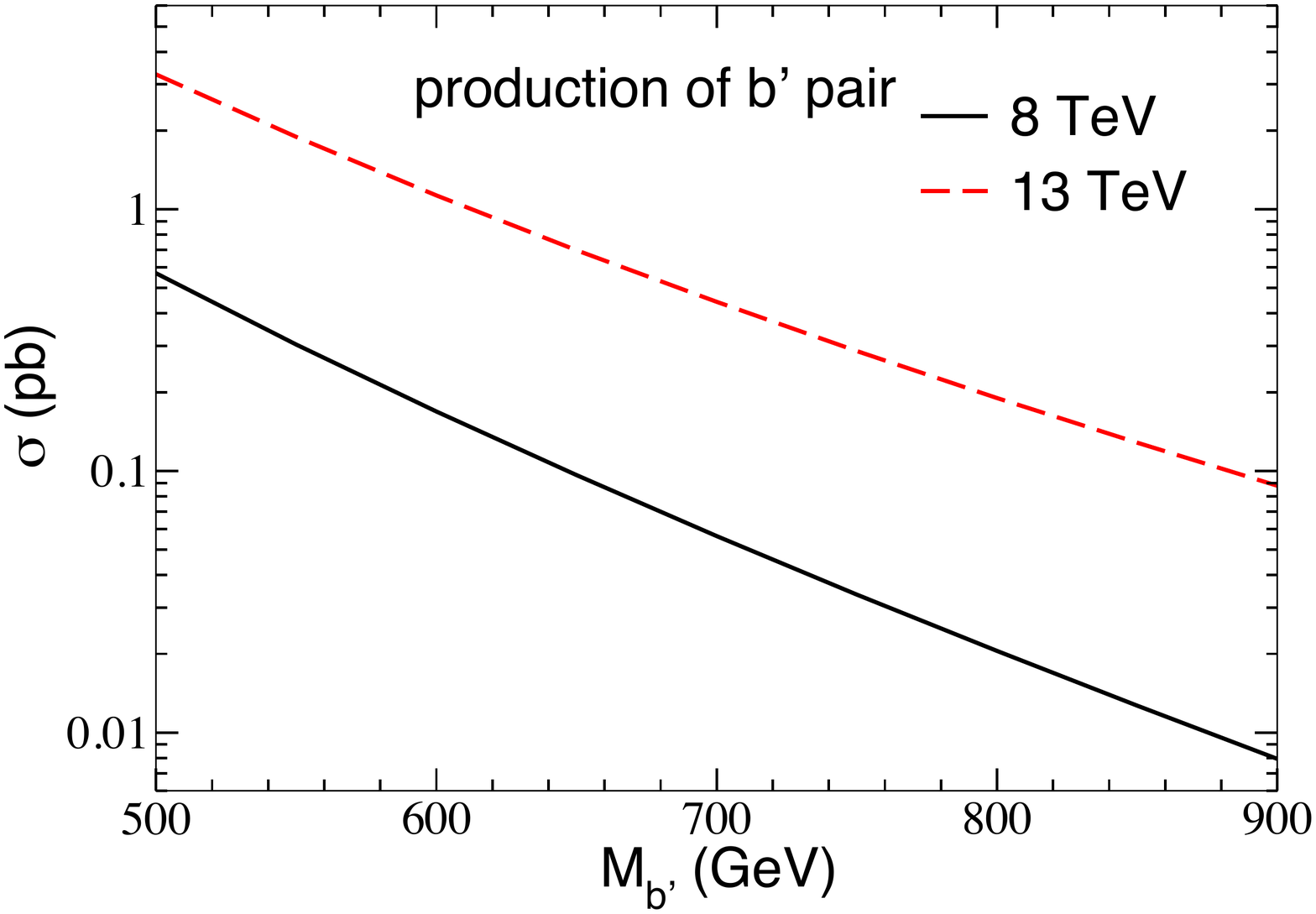}}
\subfloat[]{\includegraphics[width=3.35in, angle=0]{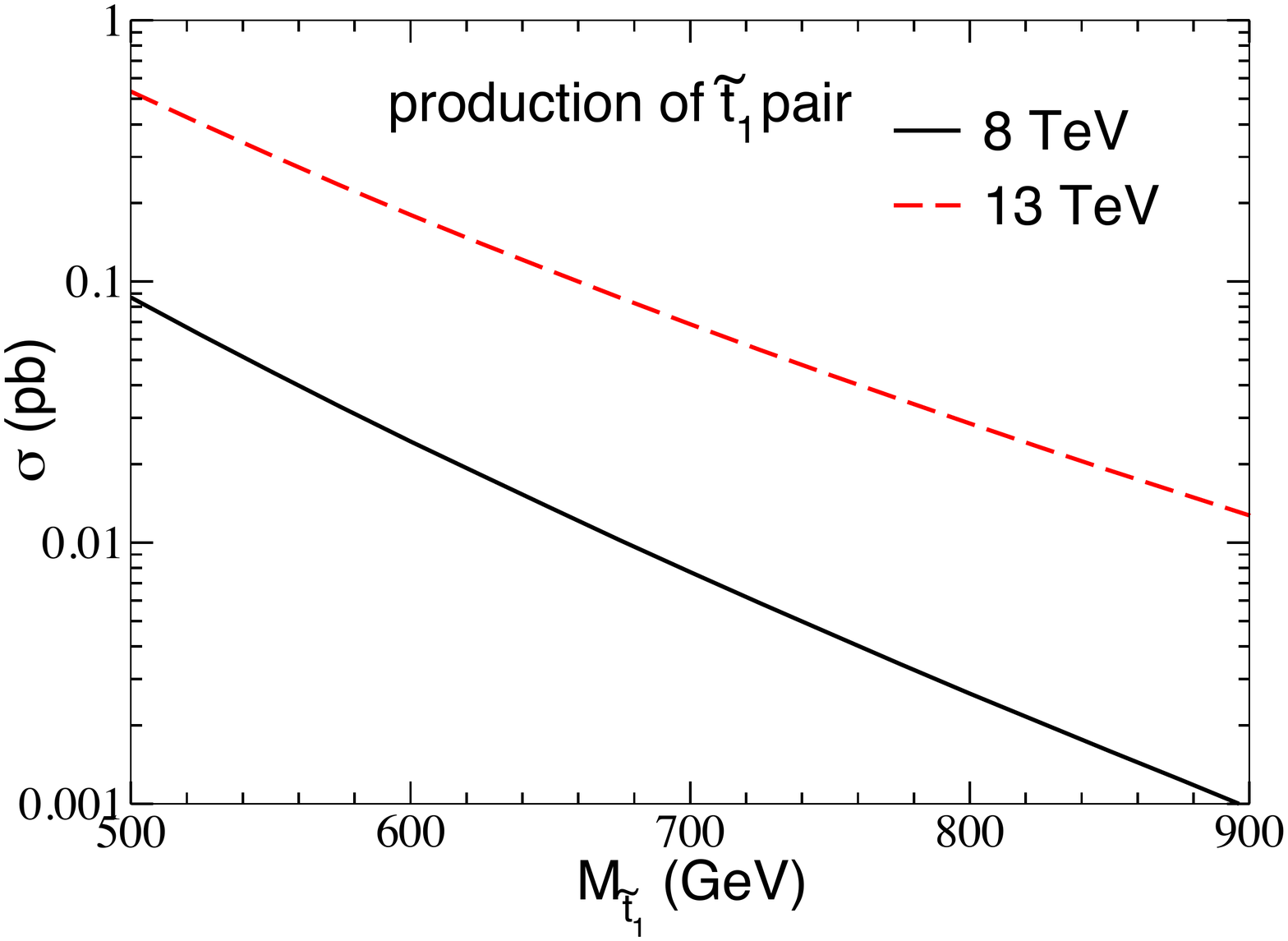}}
    \caption{\label{fig:4}\em Production cross sections for the T-odd $b^\prime$ and the stop/sbottoms in SUSY at 8 TeV and 13 TeV LHC. }
\end{figure}

It is interesting to compare with the SUSY benchmark considered in Ref.~\cite{Huang:2015fba}. There the spectrum was taken to be:
\be
m_{\tilde{t}_1} = 550 {\rm \ GeV} \ , \qquad m_{\tilde{\chi}^0_2} = 340 {\rm \ GeV} \ , \qquad m_{\tilde{\chi}_1^{\pm}} \approx m_{\tilde{\chi_1^0}} =260 {\rm \ GeV}\ .
\ee
The pair production cross section of 550 GeV $\tilde{t}_1$ is 45.2 fb at 8 TeV and the branching fraction of $\tilde{t}_1\to tW^\pm \tilde{\chi}^\mp_1$ is close to 100\%. The total signal strength is $\mu= \mu_{\tilde{t}_1} + \mu_{\rm ttH} = 2.83$. We see that the selection efficiency on the $2b+4W+E_{\rm T}^{\rm miss}$ final states for our $b'$ benchmark model is about 2.4 times the selection efficiency of the SUSY benchmark model of Ref.~\cite{Huang:2015fba}. This is because the final state particles in our benchmark model are much more energetic  due to the heavy mother particle mass as well as the large splitting in the spectrum, and hence are easier to pass the cuts.

\begin{figure}[t]
\includegraphics[width=3.35in, angle=0]{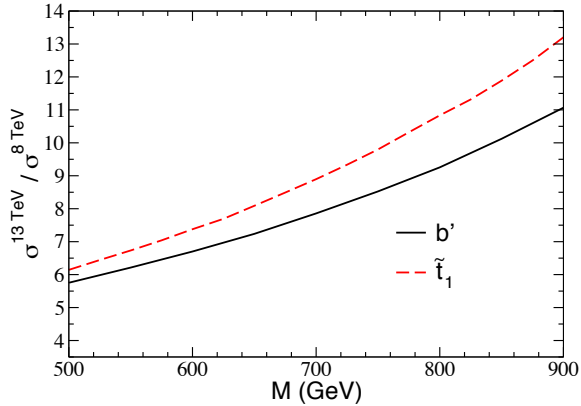}    \caption{\label{fig:4-1}\em Ratios of the cross sections for $b^\prime$ and stops/sbottoms at 13 TeV over 8 TeV. The increase in the SM ttH cross section is a factor of 3.9. }
\end{figure}

At the LHC Run 2  with the same cuts imposed in Eq.~(\ref{eq:cmscuts}), the total signal strength for the $b'$ benchmark model increases to $\mu = 3.2$. The reason of the increase is similar to the situation in the stop interpretation: the production cross-section for the heavy particles grows at a faster pace than that of SM ttH \cite{Huang:2015fba}. The increase is also more significant than that of the SUSY benchmark model [which has $\mu(13\mbox{ TeV}) =3.69$]. In Fig.~\ref{fig:4} we show the production cross sections for the T-odd $b^\prime$ and the stop/sbottom in SUSY at 8 TeV and 13 TeV LHC, while Fig.~\ref{fig:4-1} shows the ratios of the production cross sections at 13 TeV LHC over the 8 TeV LHC. In comparison, the increase in the SM ttH cross section is only a factor of 3.9 in going from 8 TeV to 13 TeV at the LHC. Fig.~\ref{fig:4} also allows for a simple scaling of cross section should one be interested in increasing (decreasing) the signal strength using a lighter (heavier) mass for the $b^\prime$ or stop/sbottoms. For example, assuming the signal acceptance stays roughly the same, a T-odd $b^\prime$ at around 620 GeV and a decay branching of 55\% into $tW_H$ could give rise to a total signal strength $\mu\approx 4$ at 8 TeV, in unit of the SM ttH signal strength. The corresponding $b^\prime$ mass for $\mu\approx 4$ in the simplified T parity model is about 720 GeV.

\begin{figure}[t]
\subfloat[]{\includegraphics[width=2.3in, angle=0]{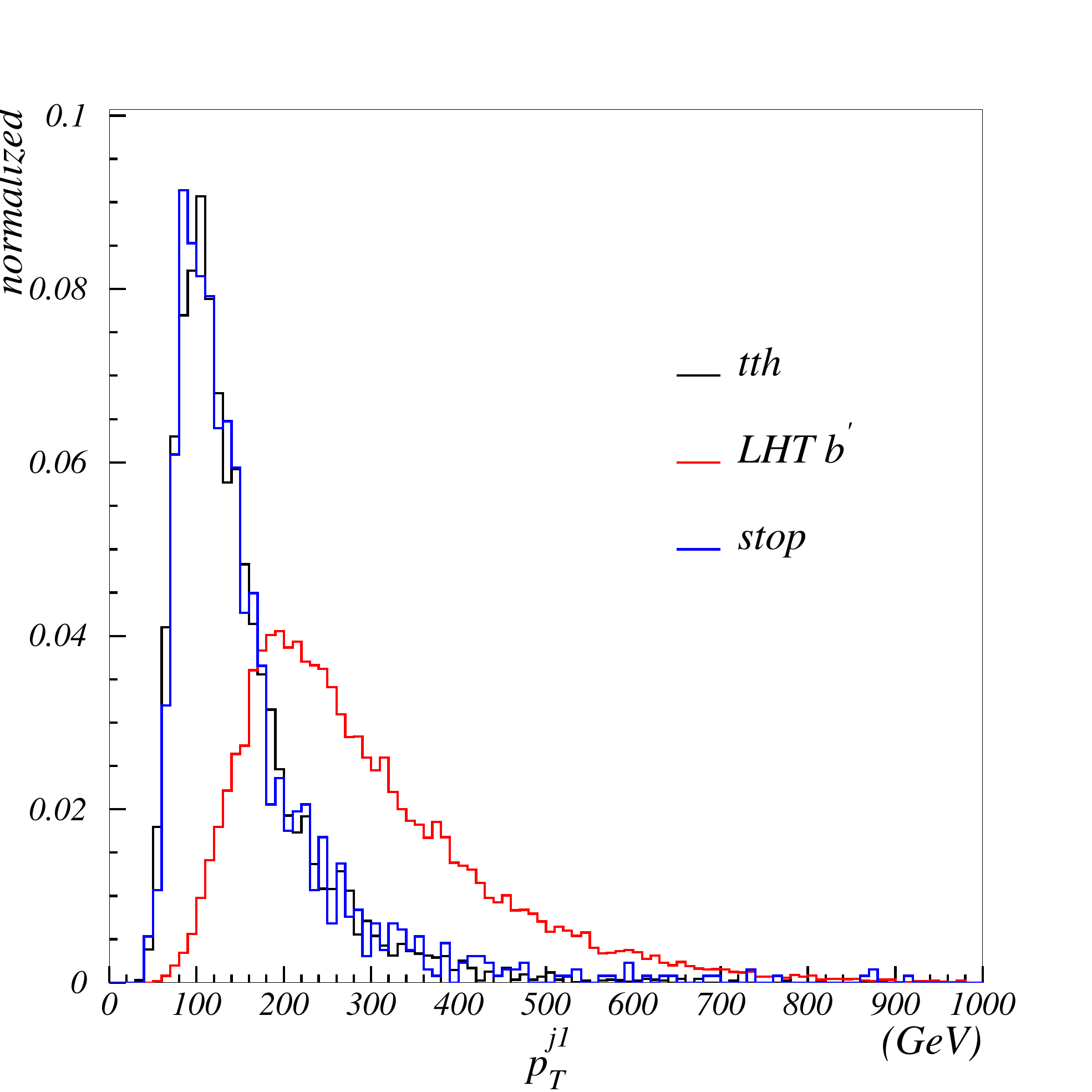}}
\subfloat[]{\includegraphics[width=2.3in, angle=0]{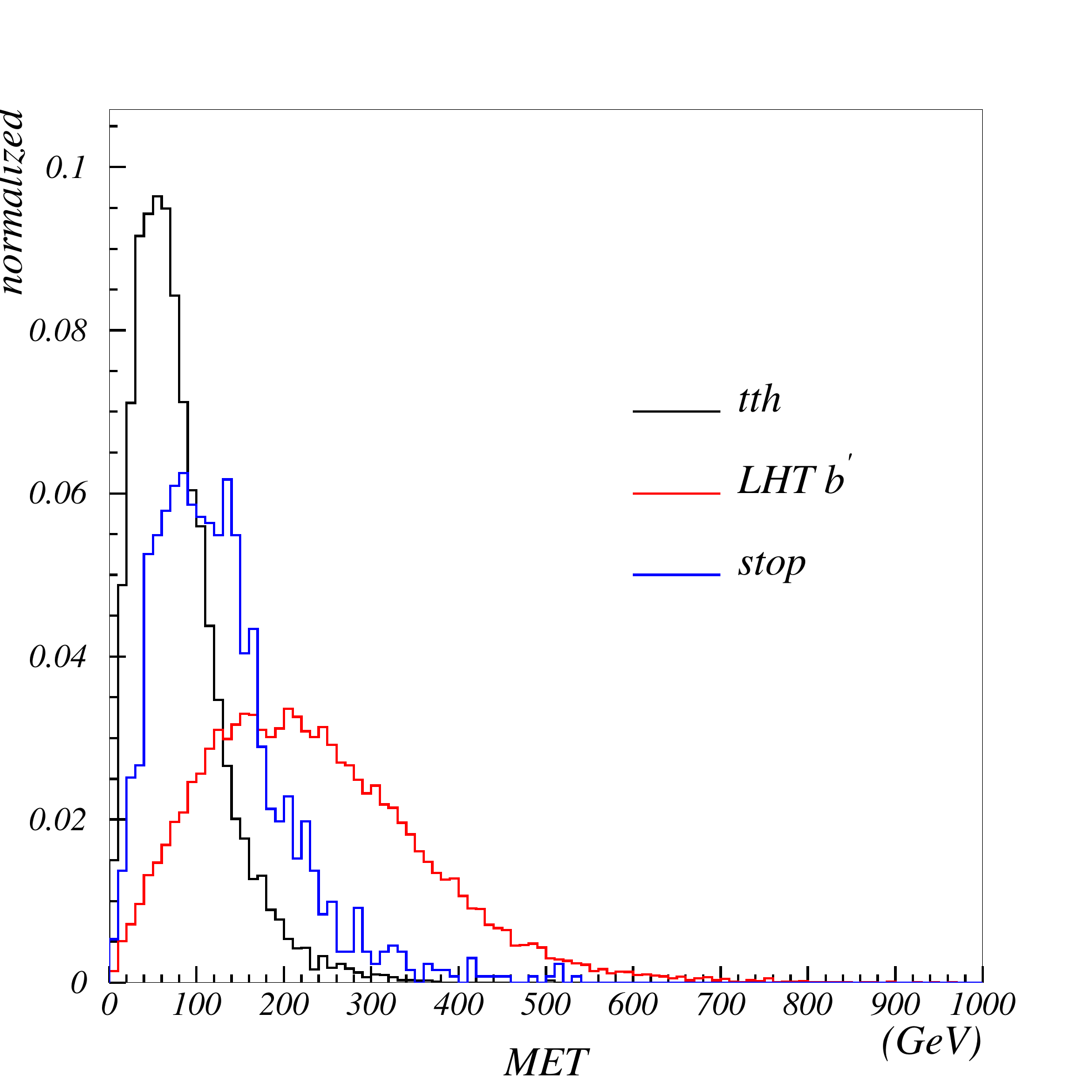}}
\subfloat[]{\includegraphics[width=2.3in, angle=0]{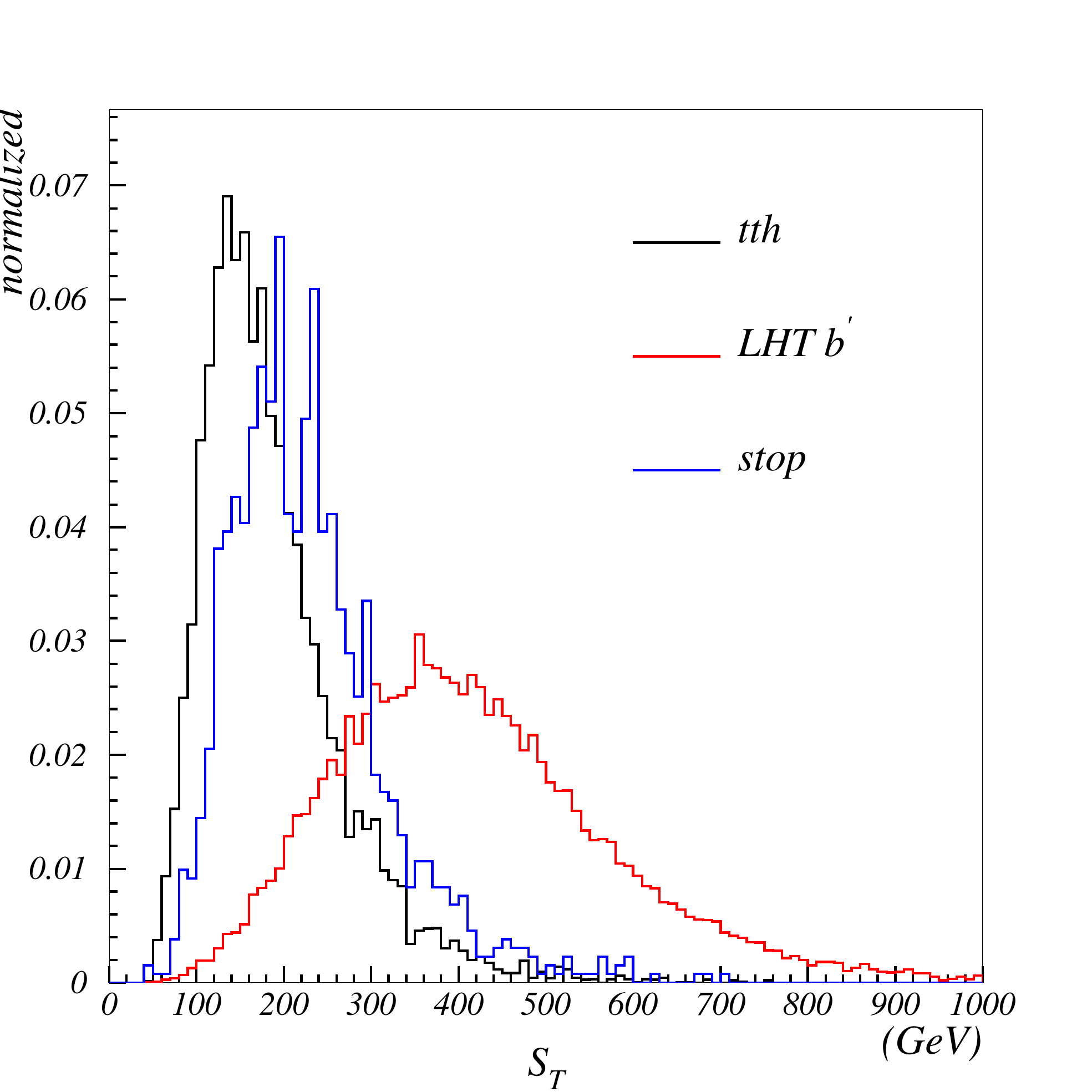}}
    \caption{\label{fig:kin}\em Normalized distributions of (a) transverse momentum the leading jet, (b) $E_{\rm T}^{\rm miss}$ and (c) $S_T=p^{\ell_1}_T+p^{\ell_2}_T+E_{\rm T}^{\rm miss}$ . The black, red and  blue histograms are from  SM ttH, T-odd $b^\prime$ with mass of $750$ GeV, and a $550$ GeV stop in MSSM considered in Ref.~\cite{Huang:2015fba}, respectively.}
\end{figure}

It will also be interesting to contemplate further kinematic cuts at the LHC Run 2 that could help disentangle the $b^\prime$ signal from that coming from the SM ttH or the stop/sbottom SUSY signals. In this regard, we show in Fig.~\ref{fig:kin} some kinematic distributions of events from the SM ttH, the T-odd $b^\prime$ and the stop in SUSY. One sees that $b^\prime$ has the hardest spectra among the three benchmarks, which is due to the heaviness of the $b^\prime$ in the benchmark, resulting in more energetic decay products. This feature is quite generic, since the fermion has a significantly larger cross section than the scalar at the same mass. Therefore, given a particular signal strength, one can always fit it with a fermion mass that is heavier than that of a scalar. As long as the spectrum is not degenerate, it will result in harder distributions of the decay products.

Based on the above observation, we can consider the following  additional cuts on the leading jet $p_T$, $E_{\rm T}^{\rm miss}$ and $S_T$,
\be
\label{eq:hardcuts}
p^{leading~jet}_T > 100 ~{\rm GeV}\ , \quad E_{\rm T}^{\rm miss} > 150~{\rm GeV}\ , \quad S_T >200~{\rm GeV},
\ee
to enhance  the contribution of $b^\prime$ to  SS2l signal region in the context of the CMS SM ttH analysis. With these further cuts, the total signal strength, in unit of the SM ttH strength, raises dramatically to
\be
\mu \approx 19 \ .
\ee
In other words, the signal would come almost exclusively from T-odd $b^\prime$.

\section{A Broader Picture}
\label{sect:broadpic}

So far we have focused on fitting the SS2l excess in the context of the ttH multilepton analyses \cite{atlastth, Khachatryan:2014qaa}, by normalizing to the SM ttH signal strength. Given that several other analyses \cite{Aad:2014pda, Aad:2015gdg, Chatrchyan:2013fea}, outside of the Higgs searches, have also observed excessive events in the SS2l category, it is worth exploring whether the T-odd $b^\prime$, or even the supersymmetric stop/sbottom, can simultaneously explain the excesses observed in these other analyses. Since these semi-independent analyses utilize different background subtraction methods and signal selection cuts, $b^\prime$ and sbottoms/stops models can have varied responses to these analyses. A detailed study on the consistency of these excesses is best carried out by the experimental collaborations. Here we will be content with some crude estimates based on publicly available information provided by the experimental collaborations, as well as our own Monte Carlo simulations. Given that the number of events in the excess of each analysis is small, it is unlikely that one can draw definite conclusions from the comparisons right now. However, it gives us a flavor on what can be done in distinguishing different models from analyses based on different selection criteria if the excesses are confirmed at 13 TeV LHC with a larger luminosity.

\subsection{CMS SUSY Analysis}

The SS2l analysis in the CMS SUSY working group in Ref.~\cite{Chatrchyan:2013fea} presented exclusion limits on the sbottom decay chain
\be
pp\to \tilde{b}_1^* \tilde{b}_1 \to (tW^-\tilde{\chi}^0_1)(\bar{t}W^+\tilde{\chi}^0_1) \label{eq:sbottom-pair}
\ee
that are degraded from the expected limits, implying more events were observed than expected. The most significant excess appeared in the signal region SR24, which requires SS2l and
\be
 N_{\rm b-jets}\ge 2 ,\quad N_{\rm jets}\ge 4, \quad 50\ {\rm GeV} \le E_{\rm T}^{\rm miss} \le 120\ {\rm GeV}, \quad H_{\rm T}\ge 400\ {\rm GeV}. 
\ee
 The expected number of events in SR24 is 4.4$\pm 1.7$ (2.8$\pm 1.2$)  in the low (high) $p_T$ region and the observed numbers are 11 (7). 
 
We simulated the contribution to SR24 from both the $b^\prime$ and the stop benchmarks. The $b^\prime$ (stop) would give rise to 0.4 (3) and 0.3 (2.3) events in the low and high $p_{\rm T}$ region, respectively. It is clear that T-odd $b^\prime$ has a lot more difficulty fitting the SR24 excess than the stop benchmark. This is due to the fact that SR24 only selects events with a relatively small $E_{\rm T}^{\rm miss}$, while the $E_{\rm T}^{\rm miss}$ distribution from $b^\prime$ benchmark is much harder than that from the stop, as can be seen from Fig.~\ref{fig:kin}.

It is also interesting to note that the $b$-tagging selection efficiency is typically 70\%. As a result, events contributing to $N_{\rm b-jets}=2$ region will also contribute to $N_{\rm b-jet}=1$ region. Therefore both the $b^\prime$ and the stop benchmarks should contribute to SR14, the $N_{\rm b-jet}=1$ cousin of SR24. However, SR14 in Ref.~\cite{Chatrchyan:2013fea} observed 6 events, which is less than the expected number of 10$\pm 4$ events. In addition, a model that could generate the excess events in the SR24 region may also produce events in other signal regions. For example, both the $b'$ and the SUSY benchmarks have significant portions of their $E_{\rm T}^{\rm miss}$ distributions beyond the upper limit 120 GeV of the SR24 region. They are expected to show up also in the signal regions SR18 and SR28 which are similar to SR14 and SR24 but requiring $E_{\rm T}^{\rm miss}>120$~GeV. The SR18 region ($N_{\rm b-jets}=1$) does have a small excess ($6.8\pm 2.5$ expected and 11 observed) but no excess ($3.4\pm 1.3$ expected and 3 observed) is seen in SR28 ($N_{\rm b-jets}\geq2$). Clearly, a consistent picture has not emerged from the current CMS SS2l SUSY search data.
Of course, we are considering a very low number statistics, of the order of five signal events or less, so we should not attempt to over-fit the current data. More data from LHC Run 2 could certainly help to clarify the situation. 

\subsection{ATLAS SUSY Analysis}

The ATLAS SS2l search in the SUSY group \cite{Aad:2014pda} also searched for sbottom decays as in Eq.~(\ref{eq:sbottom-pair}). The signal region SR1b, which requires SS2l and\footnote{There is an additional veto on SR3b signal region, which requires SS2l or three leptons, $N_{\rm b-jets}\ge 3$, $N_{\rm jets}\ge 5$ and $m_{\rm eff}> 350$ GeV.}
\be
 N_{\rm b-jets}\ge 1, \quad N_{\rm jets}\ge 3, \quad E_{\rm T}^{\rm miss}> 150\ {\rm GeV},\quad m_{\rm T}>100\ {\rm GeV}, \quad m_{\rm eff}> 700\ {\rm GeV}\ ,
\ee 
expects 4.7$\pm$2.1 events and observed 10 events with a $p$-value of 0.07, while the SR3b region expects $2.2\pm0.8$ and observed 1 event.

We implemented the above  cuts in both the $b^\prime$ and the stop benchmarks, which contributed 3.7 and 3.1 events, respectively, to the SR1b signal region. Therefore, the excess observed by the ATLAS SUSY search could be explained by both  the $b^\prime$ and the stop benchmarks. The selection efficiency for the $b^\prime$ benchmark is much better than that from the stop because of the hard cuts on $E_{\rm T}^{\rm miss}$ and $H_{\rm T}$, which can be seen from the distributions shown in Fig.~\ref{fig:kin}.

It is interesting to note that the $E_{\rm T}^{\rm miss}$ cuts in the CMS SR24 and ATLAS SR1b regions of the SUSY analyses are mutually exclusive. On the other hand, the CMS ttH analysis made public the $E_{\rm T}^{\rm miss}$ distributions of the SS2l events in the public twiki page \cite{twiki}, which exhibit broad excesses over a wide range of $E_{\rm T}^{\rm miss}$, including those overlapping with the CMS and ATLAS SUSY analyses. Again, should the excess be confirmed, the $E_{\rm T}^{\rm miss}$ distribution of the events would be a key observable to pay attention to.

\subsection{ATLAS Exotica Analysis}

The ATLAS search for heavy vector-like quarks in the SS2l + 2 $b$-jets region reported an excess in the SR4t3 signal region, which is defined by requiring SS2l and
\be
N_{\rm b-jets}=2, \quad E_{\rm T}^{\rm miss} \ge 100 \ {\rm GeV}, \quad H_{\rm T}\ge 700\ {\rm GeV} \ .
\ee
In this signal region $4.4\pm 1.1\pm 1.1$ events are expected and 12 are observed. There is also a separate excess in SR4t4,  the 3 $b$-jets category, which sees 6 events while expecting only $1.1\pm 0.9 \pm 0.4$ events. The $p$-values for both excesses are about 2$\sigma$. However, recall that the ATLAS SUSY search in Ref.~\cite{Aad:2014pda} sees no excess in SR3b, the three $b$-jets region.

Implementing the above selection cuts in our simulation, we find that the $b^\prime$ benchmark  contributes about 2.3 events to SR4t3, while the stop benchmark contributes about 1.8 events. Again $b^\prime$ contributes more to the signal region because it has a harder decay spectrum.

\section{Summary and Outlook}
\label{sect:outlook}

In this work we studied the possibility of fitting the SS2l excess observed in the LHC Run 1 data by vector-like quarks that are odd under a new parity at the TeV scale, the T parity. Phenomenology of T-odd quarks is quite different from those heavy quark decay chains that are being searched for at the LHC so far, since the T-odd quarks decays into SM particles plus the LTP, which carries away extra missing transverse energy at collider detectors.  The current bounds on the masses of the third generation T-odd quarks can be estimated by recasting the exclusion limits for stops and sbottoms in supersymmetry, which have similar decays chains to the third generation T-odd quarks,. 

We proposed a vector-like quark benchmark containing a 750 GeV $b^\prime$ quark, decaying 55\% of the time into a SM top quark and a 320 GeV $W_H$. The $W_H$ subsequently decays into the SM $W$ boson and the LTP, whose mass is at 66 GeV.  Using the ttH searches as a starting point, we normalized the 
$b^\prime$ signal strength to that of the SM ttH expectation and obtained a total signal strength of $\mu=2$ in the context of ttH searches.

At the LHC Run 2,  kinematic cuts are suggested to further enhance the $b^\prime$ signal over the SM ttH signal. Moreover, we also studied differences in the kinematic distributions between the $b^\prime$ benchmark and the stop benchmark proposed in Ref.~\cite{Huang:2015fba}, and observed that the decay spectra of $b^\prime$ are generically more energetic than those from the stop decays. Should the SS2l excess be confirmed at the LHC Run 2, it would be a top priority to determine whether the excess is due to the production of new colored particles, as well as the possible quantum numbers of such new particles.

Looking beyond the excess in the ttH searches, we also considered whether the SS2l excesses in searches for new particles can be explained in a common framework. As it stands there is some difficulty in explaining all the excesses using the $b^\prime$ or the stop benchmarks. However, the number of expected signal events is quite small at the LHC Run 1 and obviously more data from Run 2 is needed to clarify the nature of the excess. In particular, if the excess is due to the pair-production of new colored particles,  a small amount of data, of the order of 5 fb$^{-1}$, at the 13 TeV LHC is sufficient to reach the same sensitivity as the entire Run 1 dataset. 

Last but not least, even if the SS2l excess disappears as more data is collected, it is clear that searches for vector-like quarks at the LHC need to be extended to the well-motivated scenario of T-odd quarks, whose decay phenomenology requires dedicated analyses that are not covered by current searches for  vector-like quarks.

\begin{acknowledgments}
C.-R.~C. would like to acknowledge the support of National Center for Theoretical Sciences (NCTS). H.-C.~C. would like to thank Academia Sinica in Taiwan for hospitality while part of this work was done. I.~L. acknowledges helpful discussions with Aurelio Juste regarding the ATLAS exotica search. We thank the authors of Ref.~\cite{Huang:2015fba} for providing the event file for the stop benchmark used in their study. The work of C.-R.~C.  is supported in part by the National Science Council of R.O.C. under Grants No.~NSC 102-2112-M-003-001-MY3. H.-C.~C is supported in part by U.S. Department of Energy grant DE-SC-000999.  I.~L. is supported in part by the U.S. Department of Energy under Contracts No. DE-AC02-06CH11357 and No. DE-SC0010143.
\end{acknowledgments}


\end{document}